\documentclass[pra,twocolumn]{revtex4}
\usepackage{graphicx}
\usepackage{color}
\usepackage{amsmath}
\usepackage{dcolumn}
\usepackage{bm}
\usepackage{amssymb,amsmath}
\usepackage{color}

\newcommand{\bsu}{\begin{subequations}}
\newcommand{\esu}{\end{subequations}}

\newcommand{\ket}[1]{\left| #1 \right\rangle}
\newcommand{\bra}[1]{\left\langle #1 \right|}

\newcommand{\proj}[1]{| #1\rangle\!\langle #1 |}
\newcommand{\Tr}{\mathrm{Tr}}
\newcommand{\expect}[1]{\left\langle#1\right\rangle}

\newcommand{\sx}{\sigma_x}
\newcommand{\sy}{\sigma_y}
\newcommand{\sz}{\sigma_z}
\newcommand{\bro}[1]{\bar{\rho}_{#1}}
\newcommand{\rhohat}{\hat{\rho}}

\newcommand{\rhotomo}{\rhohat_{\scriptscriptstyle\mathrm{tomo}}}

\newcommand{\cL}{\mathcal{L}}
\def\Id{1\!\mathrm{l}}

\begin{document}
\title{When quantum tomography goes wrong: drift of quantum sources and other errors} 
\author{S.J.~van Enk$^{1,2}$ and Robin Blume-Kohout$^3$}
\affiliation{$^1$Department of Physics and
Oregon Center for Optics\\
University of Oregon, Eugene, OR 97403\\
$^2$ Institute for Quantum Information, California Institute of Technology, Pasadena, CA 91125\\
$^3$ Sandia National Laboratories, Albuquerque, NM 87123}

\begin{abstract}
The principle behind quantum tomography is that a large set of observations -- many samples from a ``quorum'' of distinct observables -- can all be explained satisfactorily as measurements on a \emph{single} underlying quantum state or process.  Unfortunately, this principle may not hold.  When it fails, any standard tomographic estimate should be viewed skeptically.  Here we propose a simple way to test for this kind of failure using Akaike's Information Criterion (AIC). 
We point out that the application of this criterion in a quantum context, while still powerful, is not as straightforward as it is in classical physics. This is especially the case when future observables differ from those constituting the quorum.
\end{abstract}

\maketitle

\section{Introduction}
\subsection{General remarks}
The goal of quantum-state tomography \cite{Paris2004}
is to give a statistically reliable estimate of a quantum state $\rho$.
Two further questions may come to mind: (i) what is the purpose of that estimate $\rho$? And (ii), why or when are we correct in giving an estimate of {\em just one} quantum state?

There are at least two answers to the first question: our experiment may be aimed at producing a particular state, say, a cluster state, and we may just want to verify how close $\rho$ is to the desired state. But that answer provides really just an intermediate goal. The ultimate goal is always to use the desired state for some particular quantum information processing task. So we could say that the goal of producing an estimate $\rho$ is to be able to predict the future performance in a particular protocol of one or more unmeasured quantum system(s) produced by the same source. 

Now there is a nice statistical method
for ranking different models according to their ability to predict future measurement results ({\em not} on how well they fit the past data!), based on the Akaike Information Criterion (AIC) \cite{Akaike}. That criterion was developed entirely within a classical context, but it ought to apply to quantum-state estimation, too. We show this is true, even though we will point out some interesting differences between classical and quantum statistics.

The motivation behind the second question is as follows.
Since we do not have full control over all physical quantities relevant
to the quantum-state generation process (for example, even the best laser suffers from phase diffusion; and there are always spatially and temporally fluctuating magnetic and electric fields), the quantum states produced by a quantum source are not all identical. A possible description of the individual states of $M$ systems $k=1\ldots M$ would be a sequence $\{\rho_k,k=1\ldots M\}$ where each $\rho_{k+1}$ is a little different from the previous one (even with entanglement or correlation between the different systems, we can define $\rho_k$ by tracing out all the other systems).
So, why would we use just a single estimate $\rho$ in this case? One aspect of the answer is, of course, that we have no way of estimating each individual $\rho_k$. A more positive answer is that multiple measurements of a given observable $\hat{O}$ only yield estimates of {\em average} quantities such as $\expect{O}=\overline{\Tr\rho_k\hat{O}}$ or $p_n=\overline{\Tr\rho_k\proj{O_n}}$, where the average is over those $k$ on which $\hat{O}$ was measured, and where $\ket{O_n}$ denotes an eigenstate of $\hat{O}$. These averages being linear in $\rho_k$ are determined by a {\em single} density matrix, namely the average density matrix $\rho=\overline{\rho_k}$.
This simple picture has been made much more rigorous by Renner in \cite{Renner2007}.
He showed that the crucial ingredient (missing in the simple picture) is permutation invariance. 
That is, if we randomly permute the sequence of quantum systems, and then trace out some subset, the joint state of the remaining systems is to a good approximation independently and identically distributed (i.i.d.). 
In our context this means that as long as the quorum of observables is measured in a random order, then to a good approximation any one of the remaining unmeasured systems can be described by a single density matrix $\rho$. We now discuss what may go wrong if we measure the observables constituting a quorum in a nonrandom order.
\subsection{Possible errors in standard quantum state tomography} 
It is much easier to measure a given observable from the quorum many times in a row, before switching to measurement of the next observable. Such a procedure is standard practice, but it voids Renner's proof, and so it may be that there is not a single density matrix that can be validly assigned to the remaining unmeasured quantum systems. 

Let us introduce this problem with a simple example.  Given an ensemble of $3N\gg1$ qubits that -- we assume! -- are identically and independently prepared, we want to estimate their density matrix.  So we divide them into three equal \emph{and sequential} groups, and measure $\sx$ on samples $1\ldots N$, $\sy$ on samples $N+1\ldots 2N$, and $\sz$ on the last $N$. Now, if the samples are indeed identically prepared in some state $\rho$, then we can safely perform the measurements in this order -- the state $\rho^{\otimes 3N}$ is invariant under permutations, so all orderings are equivalent.  But if the source is drifting over time, the first $N$ copies are best described by a mean density matrix $\bro{1}$, while the second and third sets of $N$ qubits are best described by (possibly different) average states $\bro{2}$ and $\bro{3}$, respectively.

For an amusing (albeit extreme) example, consider a situation where the first $N$ copies are best described by $\bro{1} = \proj{+}$, the second group by $\bro{2}=\proj{+i}$, and the third by $\bro{3}=\proj{0}$.  The measurement outcomes in this case are not random at all:  every single measurement (of $\sx$, $\sy$, $\sz$) will yield eigenvalue $+1$.  Linear inversion tomography will yield a radically non-positive state
\begin{equation}
\rhotomo = \left(\begin{array}{cc} 1 & \frac{1+i}{2} \\ \frac{1-i}{2} & 0 \end{array}\right),
\end{equation}
and maximum likelihood estimation (MLE) yields the projector onto $\rhotomo$'s positive eigenspace.  Although both estimates are plausible answers to "What single matrix best fits the observed data?", neither one of them is of any predictive use at all!  The source is drifting so rapidly and drastically that \emph{this} set of 3N samples really tells us almost nothing about future observations.  This is the simplest and best conclusion at which our data analysis should arrive.

This is a rather extreme and contrived example of experimental drift [below we will discuss a more common type of nonrandom experiment where the above cycle of measurements is repeated once: so we measure $\sx$ on the first $N/2$ copies, then $\sy$, then $\sz$, and then $\sx, \sy,\sz$ again, each on $N/2$ sequential copies].  More realistic examples show similar behavior, though.  The statistics given above are actually more consistent with a different (and still plausible) mechanism:  When the measurement apparatus is ``rotated'' to perform a different measurement, the experimenter inadvertently ``rotates'' the samples as well.  A particularly na\"ive version of this could occur with photon polarization, where one way to physically rotate a polarizer is for the experimentalist to simply rotate his own frame of reference (e.g., by lying down).  Such a passive rotation obviously fails to change the relative orientation of samples and apparatus.  More realistic examples occur when similar quantum gate devices are used to (1) prepare states (e.g. EPR states) and (2) implement measurements.  In quantum process tomography, this sort of pitfall is well known; it violates the conditions for complete positivity of processes, and causes negative eigenvalues just as in our example above \cite{WeinsteinJCP04}.  

All of these failures are examples of a single phenomenon:  \emph{sample-apparatus correlation}.  In process tomography, this is usually explained by correlation between the system and its environment.  In state tomography, there is no environment per se, but if the state of the $k$th sample is (in any way) correlated with the behavior of the measurement apparatus (e.g., with what measurement it is oriented to perform), then tomography goes wrong.  Experimental drift is a simple and easy to understand example:  the sample state is correlated with time, and if the apparatus setting is also allowed to vary with time, then there will be sample-apparatus correlation.  As noted above, this can be eliminated by explicitly randomizing the order of measurements, so that while the samples are still time-dependent, the apparatus is not.  Other kinds of sample-apparatus correlation are not so easy to remedy.

In the example given above, the extremity of the data -- and the fact that the linear inversion estimate is radically negative -- are a dead giveaway.  On the other hand, linear inversion can produce negative estimates even with ideal data \cite{BlumeKohoutNJP10,JamesPRA01} because of statistical fluctuations.  The \emph{raison d'etre} of MLE is to fix this negativity, but by constraining the estimate to positive states, MLE also hides the tell-tale signature of failed tomography.  Moreover, negative estimates are not (in general) a reliable symptom even of drastic experimental drift.  If the drifting states in the example above were a bit more mixed -- e.g. $\bro{k}' = \frac12\bro{k} + \frac14\Id$ -- then linear inversion and MLE would yield identical and positive density matrices.  But, just as in the original example, those estimates would be useless and not predictive.

Fortunately, there is a general solution to this problem.  It elegantly generalizes the observation (made above) that a radically negative $\rhotomo$ should trigger skepticism.  It can also diagnose drift in the absence of negativity if the data are sufficiently rich.  It is called \emph{model selection}.

The core principle is that, when tomography fails:
\begin{enumerate}
\item The standard model for tomography -- i.i.d. samples described by a single density matrix -- is bad.
\item Some other model will be better.
\item We can quantify ``bad'' and ``better'', and use the results to decide whether our tomography went wrong.
\end{enumerate}
Clearly, putting this into practice requires that we come up with alternative models to describe the data.  Model \emph{design} is more of an art than a science.  Here, we demonstrate alternative models for some simple and relevant problems, and leave the rich problems of general and optimal alternative-model design to future work.  Instead, we focus on model \emph{selection}, which means determining whether (i) the standard tomographic model is pretty good, or (ii) some other model (e.g. a drifting source model) is better.
\subsection{Akaike to the rescue}
To accomplish this, we propose, as we mentioned above, to use the \emph{Akaike Information Criterion} (AIC) \cite{Akaike}.  Widely used outside of physics \cite{BookAIC1,BookAIC2}, the AIC is relatively unknown within the physics community.  However, it has been applied in astrophysics \cite{AIC2007}, entanglement verification \cite{Pavel2009}, and quantum state estimation \cite{Usami,Jun4,Guta}.  Its function is to quantify (by assigning a real number) how well a given model describes the data from a given experiment. The AIC's absolute value is not meaningful, but the \emph{relative} AIC values for multiple different models have a deep and useful meaning (see following section for a more detailed discussion of the AIC, its meaning, and its derivation).  Their simplest use is to rank all the different models, and thus to identify (a) which is the best, and (b) how significantly ``worse'' the others are.

The AIC assigns a number $\Omega_k$ to each model $k$, given by \footnote{Usually an overall minus sign and an extra factor of 2 appear on the right-hand side of (\ref{AIC}), but for our purposes it is more convenient not to include those.}
\begin{equation}\label{AIC}
\Omega_k:=\ln {\cal L}_k -K_k,
\end{equation}
where ${\cal L}_k$ is likelihood of model $k$ -- or, if model $k$ has adjustable parameters (as is usually the case), the \emph{maximum} of the likelihood over all those parameters -- and $K_k$ is the number of independent model parameters used in model $k$ to 
fit the data \footnote{Actually, when the number of measurements (e.g. $3N$ in the example given) is small, there is a correction term.  The \emph{corrected AIC} (AIC$_c$) is given by $\Omega^c_k = \Omega_k - K_k(K_k+1)/(3N-K_k-1)$, with $\Omega_k$ given by \ref{AIC}. Hereafter, we will simply assume $3N\gg K_k$.}. The larger the AIC ($\Omega_k$) is, the higher the model is ranked.  While $\Omega_k$'s absolute value is meaningless, the difference $\Delta = \Omega_k-\Omega_{k'}$ represents (roughly speaking) the weight of evidence in favor of $k$ over $k'$, measured in bits.  So, for example, if we want to report a weighted average of the two models, the ratio of the weights assigned to models $k$ and $k'$ should be $w_k/w_{k'} = \exp(\Omega_k-\Omega_{k'})$.

The AIC's simple form admits a simple interpretation:  fitting the data better (higher likelihood) is good, but extra parameters are bad.  Additional parameters must justify their existence by improving the likelihood (a measure of goodness-of-fit) by at least a factor of $e$.  This helps to prevent overfitting.  Adding adjustable parameters will always improve a model's fit -- but a good fit to past data is \emph{not} a guarantee that the model will accurately predicting future measurements.  \emph{Example}:  If we measure each of $3N$ qubits, measuring $\hat{O}_j$ on qubits $j$ for $j=1\ldots 3N$, then the best possible fit to the data is to assume that each qubit $j$ just happened to be in the appropriate eigenstate of $\hat{O}_j$ so that the probability of the observed data is $\cL = 1$!  Intuitively, this ``explanation'' is absurd.  The AIC quantifies that intuition; that model requires a huge ($O(N)$) number of parameters, and the resulting penalty will overwhelm its higher likelihood, ensuring that its AIC is far worse than that of simpler models.

To apply the AIC to our example, we need an alternative model (the ``standard model'' just uses a single density matrix for all 3N qubits).  A simple alternative that describes experimental drift (as well as some other forms of sample-apparatus correlation) is to use one density matrix for each of the 3 groups of samples.  This alternative model will always fit the data at least as well, but it may use more parameters \footnote{Intuitively, the alternative model is more contrived, and this should be reflected in its ranking.  However, it is not immediately obvious how to quantify this complexity.} The AIC ranks both models, and quantifies how much better one is than the other.  We perform and analyze this calculation for our single-qubit example (where just two models are sufficient) in Section \ref{qubit}, and
address more complicated variations on this theme -- with multiple alternative models -- in Section \ref{two}.

To conclude this (long) Introduction, we note that the the appearance of maximum likelihoods in the AIC does \emph{not} imply any privileged role for MLE estimation of states or any other physical quantities.  The likelihood is a central concept in statistics, and appears in almost every method.  In the AIC, it is used specifically to quantify goodness-of-fit, and (obviously) the AIC balances this quantity against another (model complexity).  Moreover, the AIC is used only to rank different models.  There is no implicit requirement that the highest-ranked model must be chosen exclusively (in fact, a common strategy is to average over high-ranked models), and even if the ``best'' model is chosen, we remain free to analyze that model without MLE (e.g., via Bayesian averaging).

\section{Examples}

In this Section we first treat the example from the Introduction, tomography on single qubits, in more detail (Sec.~\ref{qubit}). In this example, inconsistencies can arise only when the observed average values of $\sx, \sy, \sz$ are inconsistent with each other, which in turn can only happen if the density matrix obtained by linear inversion is unphysical.
The next example, discussed in \ref{One2}, also concerns single qubits, but now measurements of $\sx,\sy,\sz$ are each repeated once. In this (experimentally more relevant) case inconsistencies can arise when two estimates of the same quantity are  statistically different. Ad-hoc methods that just consider this particularly simple type of inconsistencies work just as well as the AIC.
In the last subsection, \ref{two}, we will consider the case of multiple qubits, in which the validity of ad-hoc methods is much harder to verify, but the AIC still works in the same manner, thus showing the universality of that method.
\subsection{One qubit, part 1}\label{qubit}
We return to tomography of single qubits, where we measure $\sx$ on the first $N$ qubits, then $\sy$ on the next $N$, and $\sz$ on the last $N$ qubits.
Denote the three thusly observed averages by $X:=\expect{\sx}_{{\rm obs}}$,
$Y:=\expect{\sy}_{{\rm obs}}$, and
$Z:=\expect{\sz}_{{\rm obs}}$.
In order to calculate likelihoods, we need the frequencies of having observed spin up ($+$) and down ($-$), respectively. They are 
given in terms of these averages by
\begin{subequations}
\begin{eqnarray}
f_{\pm}^{(x)}&=&\frac{1\pm X}{2},\\
f_{\pm}^{(y)}&=&\frac{1\pm Y}{2},\\
f_{\pm}^{(z)}&=&\frac{1\pm Z}{2}.
\end{eqnarray}
\end{subequations}
A density matrix describing just the first set of $N$ measurements really uses or needs only one parameter, $X$ (the other two parameters are, obviously, not at all determined by those data). And no matter what $X$ is, there is always a perfect fit to the data. The logarithm of the (maximum) likelihood of such a density matrix is, therefore,
\begin{equation}
\ln{\cal L}^{(x)}=Nf_+^{(x)}\ln(f_+^{(x)})+Nf_-^{(x)}\ln(f_-^{(x)})
=-NH(\tfrac{1+X}{2}),
\end{equation}
with $H(.)$ the Shannon entropy.
The same story holds for the next two sets of measurements, and so there is always a perfect fit to the data when we use the ``alternative model'' with three density matrices, and that model needs three independent parameters. We conclude that the AIC assigns the following ranking to the alternative model:
\begin{equation}
\Omega_a=-N\left\{
H(\tfrac{1+X}{2})
+H(\tfrac{1+Y}{2})+H(\tfrac{1+Z}{2})\right\}-3.
\end{equation}
The performance of the ``standard model'' depends on the value of just one number. If
\begin{equation}\label{condqub}
R^2:=X^2+Y^2+Z^2\leq 1,
\end{equation}
there is a single maximum likelihood density matrix $\bar{\rho}$ (with purity $\Tr\bar{\rho}^2=(R^2+1)/2$) that describes the whole measurement perfectly, just as  the alternative model does. The standard model also needs three parameters in this case, and the maximum likelihood is also the same as for the alternative model. So, in this case there is no real difference between the two models---we could pick $\bro{1}=\bro{2}=\bro{3}=\bar{\rho}$---and we have $\Omega_s=\Omega_a$. There is no reason to reject the standard model when $R\leq 1$.

Now let us suppose that $R>1$. We have then the choice between two descriptions: 
\begin{enumerate}
\item \underline{Alternative model}: We describe
each of the three measurements by their own density matrix. The maximum likelihood estimates of those three states satisfy
\bsu
\begin{eqnarray}
\Tr\bro{1}\sx&=&X,\\
\Tr\bro{2}\sy&=&Y,\\
\Tr\bro{3}\sz&=&Z.
\end{eqnarray}
\esu
{\em Three} independent  parameters are needed for this model. ($\bro{1},\bro{2},\bro{3}$ are underdetermined, of course, but for the purpose of finding the maximum likelihood ${\cal L}_a$ the information suffices.)
\item \underline{Standard model}: We use one density matrix  to describe all three measurements together. 
The maximum likelihood estimate of that state will be {\em pure}.  There is no known method to compute it exactly, but a generally good approximation is given by
\bsu\label{a}
\begin{eqnarray}
\Tr\bro{s}\sx&=&X/R,\\
\Tr\bro{s}\sy&=&Y/R,\\
\Tr\bro{s}\sz&=&Z/R,
\end{eqnarray}
\esu
and this state's likelihood is a strict (but generally pretty tight) lower bound on the maximum likelihood for the standard model.
{\em Two} independent parameters are needed in this model \footnote{One of the authors remains bothered by the decision to separate the standard model, effectively, into two distinct models that contain (i) all the mixed states ($K=3$) and (ii) the pure states ($K=2$), respectively.  However, this protocol is specifically discussed and justified by Burnham and Anderson (\cite{BookAIC1}, section 6.9.6).  They express a similar concern, but also provide some preliminary justification for it.  So, while further thought and research seems warranted, so does this choice.}.
\end{enumerate}
The reason we end up with a {\em pure} maximum likelihood state in the standard model is that the
single matrix fitting the data perfectly lies outside the set of physical states (it has a negative eigenvalue), and the closest physical state lies {\em on} the boundary \cite{BlumeKohoutNJP10}. In the case of qubits, this means a pure state. More precisely,
if the unphysical best-fit matrix $\tilde{\rho}$ is written in its diagonal form,
$\tilde{\rho}=\sum_{k=+,-}\lambda_k\proj{\psi_k}$, with $\lambda_+>1$ and $\lambda_-<0$, then the maximum likelihood estimate would be
$\bro{s}=\proj{\psi_+}$. The latter state has the properties (\ref{a}), as can be easily verified by explicit calculation.

Thus, when $R>1$ the 
alternative model fits the data better but uses one more parameter than does the standard model.
We can calculate the maximum likelihoods analytically in each of the two models, and thus obtain the relative AIC score of the two models:
\begin{eqnarray}\label{Oms}
\Omega_s-\Omega_a=1+N\sum_{M=X,Y,Z}
\frac{1}{2}\ln
\frac{1-M^2/R^2}{1-M^2}
+\nonumber\\
\frac{M}{2}\ln\frac{(R+M)(1-M)}{(R-M)(1+M)}.
\end{eqnarray}
We accept the standard model as consistent iff $\Omega_s\geq\Omega_a$. This will happen only if $R$ is sufficiently close to 1. If we expand $R$ around 1,  we can Taylor expand the right-hand side of (\ref{Oms}) as
\begin{equation}
\Omega_s-\Omega_a\approx  1-N\sum_{M=X,Y,Z}\frac{(R-1)^2M^2}{2(1-M^2)},
\end{equation}
provided $(R-1)^2\ll (1-M^2)$ for $M=X,Y,Z$. That is, with this proviso, the standard model is consistent only when
\begin{equation}\label{cond}
(R-1)\leq 
\frac{C}{\sqrt{N}},
\end{equation}
with the constant $C$ given by
\begin{equation}
C=\frac{1}{\sqrt{\sum_M M^2/2(1-M^2)}}.
\end{equation}
The dependence of the condition (\ref{cond}) on $N$ agrees with the simple idea that it is sufficient for $R$ to be less than about a standard deviation or two above 1 for the standard model to still apply, and that standard deviation, of course,
decays like $1/\sqrt{N}$ for $N\rightarrow\infty$.

\subsection{One qubit, part 2}\label{One2}
The implementation of tomography in the previous example is probably too simple and too obviously wrong for it to have been applied in an actual experiment. The straightforward improvement to measure each of $\sx, \sy, \sz$ in two separate blocks will allow one to detect drift. Let us denote the 6 observed averages by $X_{1,2}:=\expect{\sx}_{{\rm obs 1,2}}$,
$Y_{1,2}:=\expect{\sy}_{{\rm obs 1,2}}$, and
$Z_{1,2}:=\expect{\sz}_{{\rm obs 1,2}}$. Drift can be detected by comparing the pairs of estimates $X_{1,2}$ with each other, $Y_{1,2}$ with each other, and $Z_{1,2}$ with each other.
The AIC works as follows:
We need again at least two different models for describing the data. One will be the standard model, with one density matrix describing all 6 measurements. This density matrix will be determined by the three averages $(X_1+X_2)/2$ etc.
The alternative model may consist of two independent density matrices (with 6 parameters in total) or of two density matrices that are not independent with either 4 or 5 parameters in total. Let us test the AIC in a simulation of data generated by single-qubit states of the form
\begin{equation}
\rho_{{\rm actual}}=p\proj{\psi_\phi}
+(1-p)\openone/2,
\end{equation}
where the pure state $\ket{\psi_\phi}$ depends on an angle $\phi$, which we assume to undergo a random walk, and with the following meaning:
\begin{eqnarray}
\bra{\psi_\phi}\sx\ket{\psi_\phi}&=&\cos\phi;\nonumber\\
\bra{\psi_\phi}\sy\ket{\psi_\phi}&=&\sin\phi;\nonumber\\
\bra{\psi_\phi}\sz\ket{\psi_\phi}&=&0.
\end{eqnarray}
For $p$ we take the value $p=0.9$.
We perform in total 3000 measurements, divided into 6 groups of 500, in which we measure $\sx,\sy,\sz,\sx,\sy,\sz$ in that order.
\begin{figure}[h]
  \begin{center}
\includegraphics[width=3in]{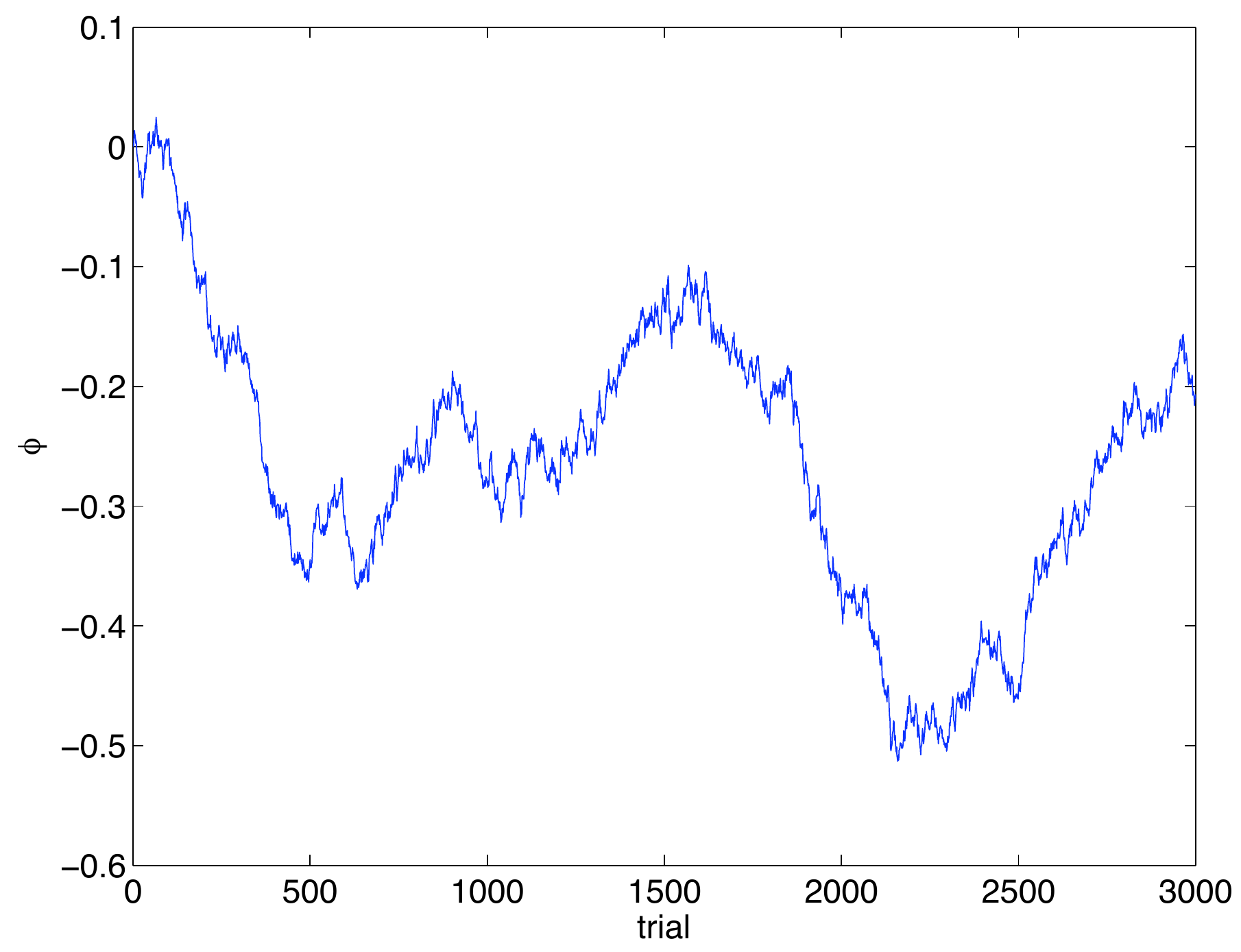}
  \includegraphics[width=3in]{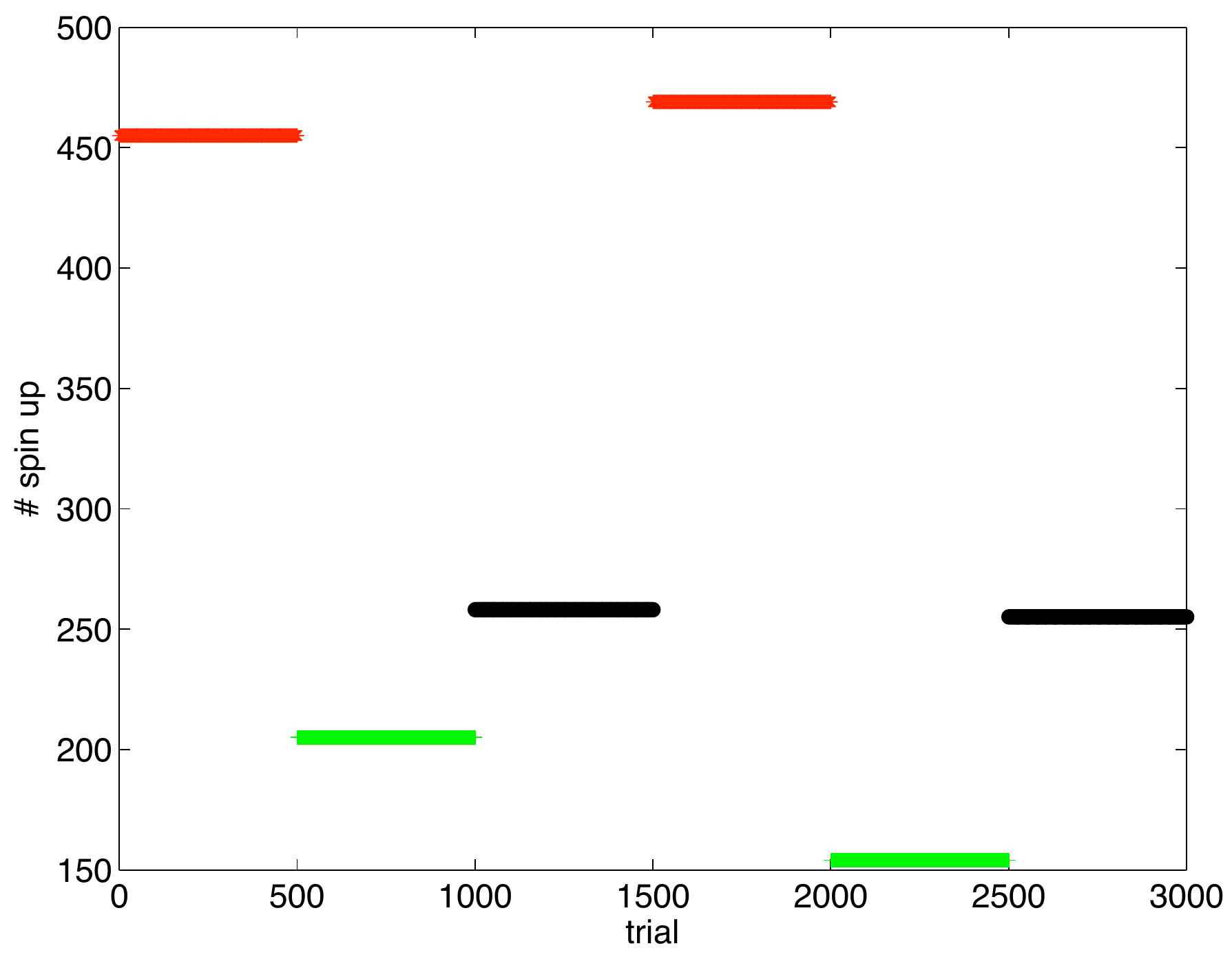}
  \caption{Top: 
  Simulation of diffusion of the angle $\phi$ in the state $\ket{\psi_\phi}$ over the course of 3000 measurements.
  Bottom: the number of ``spin up'' results for the measurements of $\sx$ (in red, for measurements 1...500 and 1501...2000), of $\sy$ (in green, for measurements 501...1000, and 2001...2500), and of $\sz$ (in black, for the remaining measurements).
  The numbers for $\sy$ are statistically different, and this is reflected in the relative ranking the AIC accords to the different models. Here we have
  $\Omega_s-\Omega_a=-5.07$, where the negative sign implies the standard model of a single density matrix is significantly worse than the alternative model (here, two density matrices with five parameters in total, two parameters more than the standard model). Tomography failed in this case.
}
  \end{center}
 \end{figure}

\begin{figure}[h]
  \begin{center}
\includegraphics[width=3in]{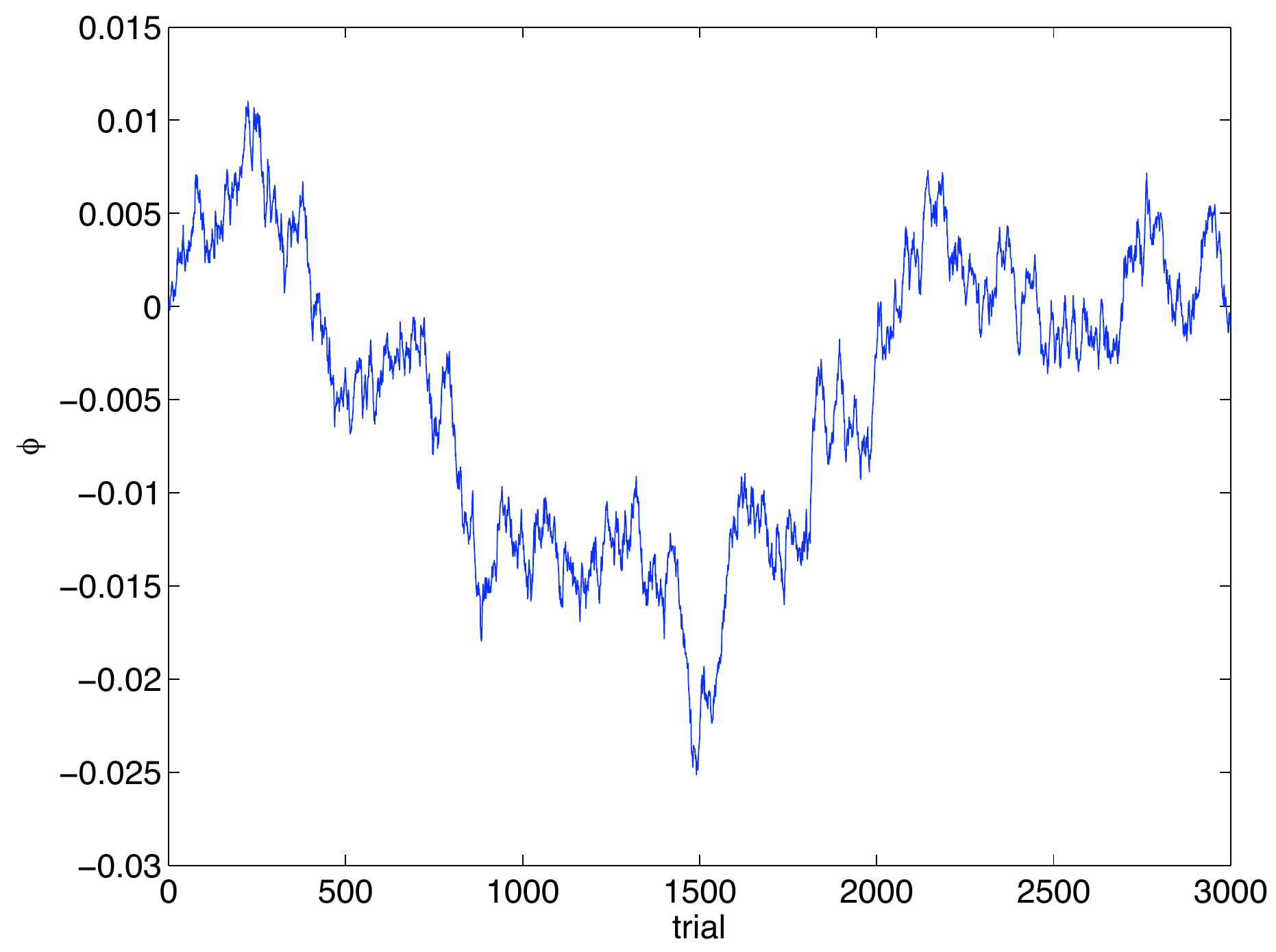}
   \includegraphics[width=3in]{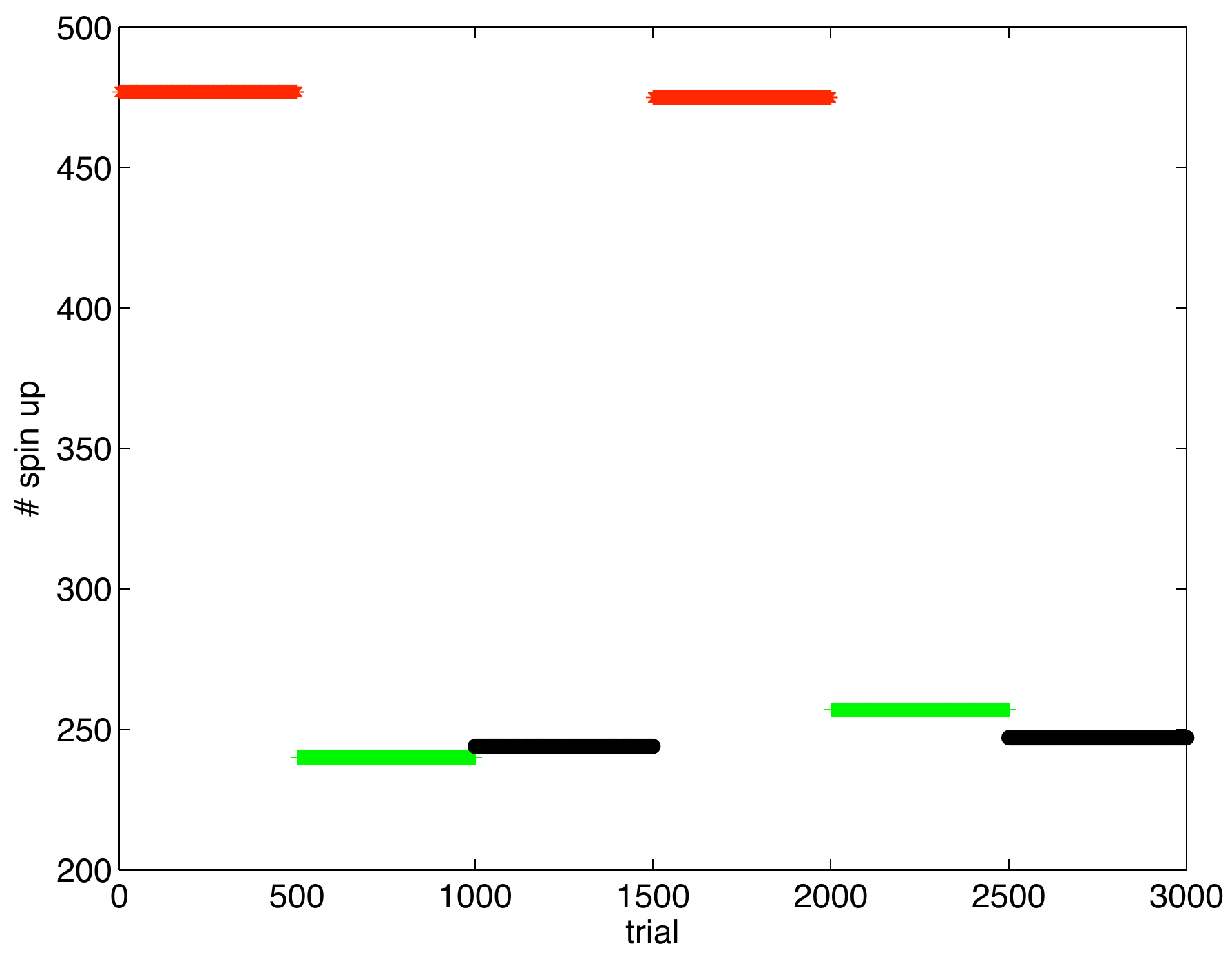}
  \caption{Same as Fig.~1, but for a case where the drift is much smaller over the course of 3000 measurements. Here
  $\Omega_s-\Omega_a=1.38$, so that the 
  standard model is better than the best alternative model (which has two extra parameters). Tomography succeeded.}
  \end{center}
 \end{figure}

In Figs 1 and 2 we plot two qualitatively different cases. In the first case the diffusion of $\phi$ is so fast that it leads to noticeably different values of $X_{1,2}$ and $Y_{1,2}$. The AIC in this case gives a very clear preference for the alternative model of using two density matrices with 5 parameters in total (only the expectation value of $\sz$ does not change over the course of the experiment). 

In the second case the drift over the course of the experiment is small enough so that the standard model is still the best, even though there is some drift, and even though the more complicated model does, of course, fit the data slightly better.

\subsection{Two or more qubits}\label{two}
Consider now a tomographically complete measurement on $9N$ copies of two qubits, where on the first $N$ pairs of qubits we measure 
$\sx$ on both qubits independently, then on the next $N$ pairs we measure $\sx$ on one qubit and $\sy$ on the other, then on the third set of $N$ pairs we measure $\sx$ on the one and $\sz$ on the other $\ldots$ until on the last (9$^{{\rm th}}$) set of $N$ pairs we measure $\sz$ on both qubits. The first measurement is described by {\em three} independent averages that are obtained from measuring $\sx$ on both qubits independently: 
\bsu
\begin{eqnarray}
XX&:=&\expect{\sx\otimes\sx}_{{\rm obs,1}},\\
IX&:=&\expect{\openone\otimes\sx}_{{\rm obs, 1}},\\
XI&:=&\expect{\sx\otimes\openone}_{{\rm obs, 1}}.\label{XI1}
\end{eqnarray}
\esu
Thus, a two-qubit density matrix
perfectly fitting the data of the first measurements needs three parameters. The description of the second measurement of
$\sx$ on one qubit and $\sy$ on the other is likewise determined by three observed averages
\bsu
\begin{eqnarray}
XY&:=&\expect{\sx\otimes\sy}_{{\rm obs, 2}},\\
IY&:=&\expect{\openone\otimes\sy}_{{\rm obs, 2}},\\
XI'&:=&\expect{\sx\otimes\openone}_{{\rm obs, 2}}.\label{XI2}
\end{eqnarray}
\esu
The new feature arising here is that we get a second estimate of the same parameter, $XI$ in this case. That is, if there were only a single two-qubit state  in the experiment, the estimates (\ref{XI1}) and (\ref{XI2}) would have to agree (within error bars). Conversely, if they do not agree, we have encountered a new diagnosis of inconsistent tomography.

Writing down all different averages
obtained from this particular experiment, we find
9 quantities that are measured once, and 6 other quantities that are measured thrice.
It becomes now much harder to judge when all the differences between those different estimates of the same quantities  are, in total, statistically significant or not. That is, the generalization of the ad-hoc method that worked fine for a single qubit, becomes troublesome. This, of course, becomes exponentially worse for more than two qubits.

On the other hand, the AIC can be applied straightforwardly to various alternative models. It is sufficient to find just one alternative model superior to the standard model in order to 
have succeeded in diagnosing an inconsistency in our tomographic experiment. Of course there is a large multitude of alternative models, but one can be guided in searching for such models by looking for those estimates of the same quantities that are the least consistent.

\section{Model selection, the AIC, and quantum quirks}


Data are generally assumed to be generated by some stochastic process \footnote{A semi-philosophical note:  it's not necessary to invoke intrinsic randomness.  The underlying process might be deterministic but so complex that its description is hopeless or just not worth the time.  The Bayesian view of probabilities is entirely compatible with this view, and permits us to \emph{describe} data and the processes that generate them without needing to ``believe in'' randomness.} -- e.g., a probability distribution $f(x)$ (where $x$ denotes the sample space containing all possible events).  Unfortunately, these ``true'' probabilities are unknown to us.  All we have are some data.  So, in order to (i) describe the data; (ii) approximate the underlying process $f$; and (iii) most importantly, predict \emph{future} observations, we use \emph{models}.

A model is just another probability distribution $g(x)$.  Almost always, the model contains a whole family of \emph{parameterized} distributions $g_\theta(x)$, where $\theta$ comprises the values of $K$ distinct [real-valued] parameters.  One obvious model is the universal one where each of the probabilities $g(x)$ -- for every possible value of $x$ -- is itself a free parameter.  This is the richest possible model, with the most parameters.  If $x$ takes on uncountably many values, this model is utterly intractable (and the AIC penalizes it infinitely for its richness).  The ubiquity of this problem in statistics motivates the use of restricted parameterized models (e.g., Gaussian distributions) where finitely many parameters can specify $g(x)$ for every possible $x$.

Quantum tomography applications usually involve finitely many parameters, but few-parameter models are still important.  This is partly because of the simplification obtained by eliminating many parameters (e.g., when a quantum state in $2^N$ dimensions is approximated by a matrix product state with $\mathrm{poly}(N)$ parameters), but even more importantly because it guards against overfitting.  This is precisely where well-designed model selection techniques come in, and the AIC is a canonical example.  When there is a choice between different candidate models describing one and the same experiment, the AIC provides a numerical ranking of the different models.

The AIC (as given in Eq.~(\ref{AIC})) appears very simple.  Moreover, it bears a strong resemblance to quantities that appear in likelihood-ratio (LR) hypothesis testing (see, e.g.~\cite{BlumeKohoutPRL10}).  But in fact, the AIC's theoretical underpinnings are rather different, and remarkably elegant (see \cite{BookAIC1} for extensive discussion).  Likelihood ratios are a fundamentally frequentist technique:  given two competing models, we calculate ahead of time the probability that various values of the LR statistic will be observed \emph{if} one model or the other is ``correct'', and then we formulate a rule for what to announce upon seeing any given value of the LR statistic.  Many canonical results on LR tests require that the models be nested -- i.e., that one be a subset of the other.  In particular, given this and a few other conditions, it is possible to derive expectation values of the LR statistic that look identical to Eq.~(\ref{AIC}) because the loglikelihood ratio is $\chi^2_K$ distributed, and has mean value $K$.

But despite this similarity, the AIC is derived differently.  Akaike began by postulating that ``goodness'' of a model is quantified by the Kullback-Leibler divergence \cite{KL51} between the model and the ``true model'' that actually underlies the data.  Then, rather remarkably, he showed that it is possible to estimate this divergence \footnote{More precisely, this estimate is in fact determined only up to a constant, but that constant is the same for all models, and hence drops out when comparing different models.} -- \emph{even when the true model is unknown!} The AIC is the expected value of the [unknown] Kullback-Leibler divergence between a specified model and the [unknown] true model, conditional upon the data in our possession.  So the AIC (i) has a powerful and universal interpretation, and (ii) can be used to compare arbitrary models, without any requirement for nesting.

This is not to say that the AIC is the acme of model selection, nor that it is perfectly adapted to quantum tomography problems.  First, there are competing derivations of other model ranking statistics, such as the Bayesian Information Criterion (BIC -- again, see \cite{BookAIC1}).  Moreover, the AIC is inherently an asymptotic result -- much like, for example, the efficiency of MLE.  So, even though there is a finite sample size correction (the AIC$_c$), this correction is part of an asymptotic expansion and may be unreliable for any fixed $N$.

One significant consequence of this is that, for finite samples, an event $x$ whose true probability is nonzero may not be observed -- in which case a model might assign zero probability to it.  (The MLE within the full model, where each probability $g(x)$ is a parameter, behaves this way).  This results inevitably in an \emph{infinite} Kullback-Leibler divergence.  Asymptotically, the probability of such a pathology occurring goes to zero almost certainly.  But for any finite sample size it is a concern. So, beware of rank-deficient estimates in tomography!

A related phenomenon is [almost] unique to quantum tomography.  Akaike's derivation assumes that a very good (if not the best) measure of predictive power is the Kullback-Leibler divergence between the true model $f(x)$ for the observed process $x$ and the assigned model $g(x)$.  But in quantum tomography, the observed process ``$x$'' is some particular (and rather arbitrary) quorum of measurements that the tomographer has performed.  We don't necessarily care about predicting \emph{those} measurements!  Instead, we care about the underlying quantum state -- or, to put it more operationally, we care about a large and unknown set of \emph{other} measurements that might be performed on samples of that state in the future.  Quite frequently, we care about measurements of that state's diagonal basis.  This completely undermines Akaike's assumption (that predicting $x$ is the goal).  This does not mean that the AIC should not be used -- but it does strongly suggest that:
\begin{enumerate}
\item Conclusions drawn from the AIC, or any other classical statistical method, should be treated with thoughtful care,
\item Better methods may still be derived (e.g., a ``quantum AIC'')
\item Estimates obtained via the AIC should \emph{not} be expected to have good properties with respect to quantum relative entropy (the quantum version of Kullback-Leibler divergence).
\end{enumerate}
Importantly, however, there are cases where our future measurements will be the same as those used for our preliminary quantum tomography experiment. For instance, in the case of quantum computing, where error correction is implemented by CSS codes, all measurements will be Pauli measurements. In such a case, the conclusions of the AIC, applied to a tomography experiment that used Pauli measurements as well, should be trustworthy.

\section{Summary and Discussion}\label{discuss}

Our central message here is that when the assumptions of tomography fail, it is often due to some sort of sample-apparatus correlation, and that this can be detected with statistical reliability by model selection using the AIC.  One particular example, the drifting source, clearly voids the single-density-matrix model, but can be described naturally (and more accurately!) by multiple density matrices associated with different times and/or measurement settings.  The AIC is a particularly good and elegant tool for identifying whether the added complexity of this model is justified.  Ultimately, the point of model selection (especially using the AIC) is to get better predictions of future measurement outcomes -- \emph{not} just better fits to observed data.

While the AIC ranks competing models, by assigning each model $k$ a number $\Omega_k$, through Eq.~(\ref{AIC}), we have great flexibility in what to do with that ranking.  Small differences in AIC are not significant; if $|\Omega_k - \Omega_{k'}| << 1$, then both models are equally good.  But even when significant differences exist, we may choose to use the ``best'' model exclusively, or to hedge by mixing it with lower-ranked models (with weights determined by their respective AICs).  We could apply Bayesian methods to the highest-ranked model, or use maximum likelihood estimates to choose model parameters.  Choosing between these alternatives is beyond the scope of this paper.

If a model-selection (e.g., AIC) analysis finds overwhelming evidence of sample-apparatus correlation (e.g. source drift), it is often possible to go beyond the conclusion ``tomography has failed!''  What has really failed is the i.i.d. assumption -- we have convincing evidence that the samples are not identically distributed.  The joint state is therefore not (with high confidence) of De Finetti form (see \cite{Caves2002}).  But it may be possible to assign states with a relaxed De Finetti form, and thereafter to do tomography with this in mind.  For example, if the AIC declares the alternative three-state model much superior to the single-state model, one could assign a state of the form
\begin{equation}\label{DF}
\rho^{(3N)}=
\int d\bro{1}\int d\bro{2}
\int d\bro{3}\, P_a(\bro{1},\bro{2},\bro{3})\,\bro{1}^{\otimes N}
\otimes
\bro{2}^{\otimes N}
\otimes
\bro{3}^{\otimes N}
\end{equation}
to the $3N$ qubits, where $P_a(.,.,.)$ is a joint probability distribution over three 2D density matrices.  This form itself needs to be tested and validated, by comparison to a richer model (e.g., a model with 6, 9, or more different states).  In general, validating a model requires more sophisticated model design -- e.g., to describe more arbitrary forms of source drift -- and perhaps different measurements or experiments specifically aimed at detecting those models, as proposed in \cite{Lucia}.  But once a given model \emph{is} validated, if it implies a relaxed De Finetti form as in Eq.~(\ref{DF}), then we can in principle perform tomography independently on each of the i.i.d. subsets of the whole sample.

In the simplest case of tomography on single qubits, we discussed two competing models. Either one uses just a single density matrix $\bar{\rho}$ to describe the experiment [the standard model], or one uses three--$\bro{1},\bro{2},\bro{3}$--one for each set of $N$ qubits used to measure $\sx$, $\sy$, and $\sz$, respectively [the alternative model].
But what does it mean to use {\em three} density matrices for predicting future measurement outcomes?
The answer is that the predictions
refer to measurements on qubits that have not been measured yet (of course). Consider one unmeasured qubit  taken from, say, a set of $N+n$ qubits, from which $N$ qubits were randomly picked to be measured in the $\sx$ basis and $n$ were not measured. In this case, those $n$ qubits  would be assigned a state of the form
\begin{equation}
\rho^{(n)}=\int d\bro{1}\int d\bro{2}\int d\bro{3}\,
P_a(\bro{1},\bro{2},\bro{3})\,\bro{1}^{\otimes n},
\end{equation}
valid for any $n$, including $n=1$.	
The mixed model, as mentioned above, would combine the standard and alternative models and assign an even more mixed state. For example, in the case $n=1$ it would assign the  estimate
\begin{eqnarray}
\rho_{{\rm mixed}}&=&w_a\int d\bro{1}\int d\bro{2}\int d\bro{3}\,
P_a(\bro{1},\bro{2},\bro{3})\,\bro{1}\,\nonumber\\
&+&w_s\int d\bar{\rho}\,P_s(\bar{\rho})\,\bar{\rho},
\end{eqnarray}
with $w_a=\exp(\Omega_a)/(\exp(\Omega_a)+\exp(\Omega_s))$ and $w_s=1-w_a$ the relative weights of the two models, as assigned by the AIC, and with $P_s(.)$ the  standard De Finetti probability distribution over single density matrices.

Although we have avoided discussion of model design here, one simple but powerful technique deserves mention.  In the example at the beginning of the paper, we introduced an alternative model wherein each measurement setting is associated with a different density matrix.  When the measurements are informationally complete, this alternative model has precisely as many parameters as the standard model.  But if they are \emph{overcomplete}, then the alternative model has more parameters.  As long as the samples really are i.i.d., we expect the alternative model to fit slightly better, and the AIC to declare them (on average) equally good.  However, in the presence of experimental drift, we will find inconsistencies \emph{within} the overcomplete measurement set -- i.e., we will \emph{not} be able to fit all the measurements well with a single density matrix!  This is a simple test for experimental drift that does not rely on negativity of $\rhotomo$.

For the main point of this paper, however, all these complications are unnecessary.  All that matters is whether assigning a single density matrix to our tomography experiment constitutes the best model or not.  If not, something is amiss, but at least we have diagnosed the problem.

The main issue we left open is the following: is there a sense in which the AIC works reliably if future measurements are different than those used in our tomography experiment?
If not, is there a ``quantum'' version of the AIC that, e.g., takes into account the quorum of observables that have been measured, as well as the set of observables that will be measured?

(Upon completion of this paper  Ref.~\cite{Langford} appeared, which
is similar in spirit to our paper, but which uses $\chi^2$ tests to detect errors in tomography.
It points out, too, the problem with pure-state assignments for those tests.)

(After submission of the page proofs we became aware of two more relevant papers: \cite{Rosset2012,Moroder2012}.)

\section*{Acknowledgments}
 This work was supported by NSF Grant No.~PHY-1004219.  Sandia National Laboratories is a multi-program laboratory operated by Sandia Corporation, a wholly owned subsidiary of Lockheed Martin Corporation, for the U.S. Department of EnergyÕs National Nuclear Security Administration under contract DE-AC04-94AL85000.

\bibliography{wrongtomo-v3}
\end{document}